\documentclass[longauth]{aa}
\usepackage{natbib}
\usepackage{siunitx}
\usepackage{txfonts}
\usepackage[utf8]{inputenc}
\usepackage{graphicx}
\usepackage{float}
\usepackage{longtable}
\usepackage{cleveref}
\usepackage{hyperref}
\usepackage{url}
\begin{document}

\title{SEPIA -- a New Single Pixel Receiver at the APEX Telescope}


\author{V. Belitsky    \inst{1,2} \and
        I. Lapkin      \inst{1,2} \and
        M. Fredrixon   \inst{1,2} \and
        D. Meledin     \inst{1,2} \and
        E. Sundin      \inst{1,2} \and
        B. Billade     \inst{2} \and
        S.-E. Ferm     \inst{1,2} \and
        A. Pavolotsky  \inst{1,2} \and
        H. Rashid      \inst{1} \and
        M. Strandberg  \inst{1,2} \and
        V. Desmaris    \inst{1} \and
        A. Ermakov     \inst{1} \and
        S. Krause      \inst{1} \and
        M. Olberg      \inst{2} \and
        P. Aghdam      \inst{1,2} \and
        S. Shafiee     \inst{1,2} \and
        P. Bergman     \inst{2} \and
        E. De Beck     \inst{2} \and
        H. Olofsson    \inst{2} \and
        J. Conway      \inst{2} \and
        C. De Breuck   \inst{3} \and
        K. Immer       \inst{3} \and
        P. Yagoubov    \inst{3} \and
        F.M. Montenegro-Montes \inst{4} \and
        K. Torstensson \inst{4} \and
        J.-P. P\'erez-Beaupuits \inst{4} \and
        T. Klein       \inst{4} \and
        W. Boland      \inst{5} \and
        A. M. Baryshev \inst{5} \and
        R. Hesper      \inst{5} \and
        J. Barkhof     \inst{5} \and
        J. Adema       \inst{5} \and
        M. E. Bekema   \inst{5} \and
        A. Koops       \inst{5}
}

\institute{
  Group for Advanced Receiver Development (GARD),
  Department of Space, Earth and Environment,
  Chalmers University of Technology,
  41296~Gothenburg, Sweden\\
\email{\url{victor.belitsky@chalmers.se}}
\and 
  Department of Space, Earth and Environment,
  Chalmers University of Technology,
  Onsala Space Observatory (OSO),
  43992~Onsala, Sweden
\and 
  European Southern Observatory (ESO),
  Karl-Schwarzschild-Str.~2,
  85748~Garching bei München, Germany
\and 
  European Southern Observatory,
  Alonso de C\'ordova 3107,
  Vitacura, Casilla 19001,
  Santiago de Chile, Chile
\and 
  Netherlands Research School for Astronomy (NOVA),
  Kapteyn Astronomical Institute,
  Landleven 12,
  9747 AD Groningen,
  The Netherlands
}

\date{Received; accepted}
\abstract
{We describe the new SEPIA (Swedish-ESO PI Instrument for APEX)
  receiver, which was designed and built by GARD OSO in collaboration with
  ESO. It was installed and commissioned at the APEX telescope during 2015
  with an ALMA Band~5 receiver channel and updated with a new frequency
  channel (ALMA Band~9) in February 2016.}
{This manuscript provides a reference for observers who use the SEPIA receiver
  in terms of the hardware description, optics and performance as well
  as the commissioning results.}
{Out of three available receiver cartridge positions in SEPIA, the
  two current frequency channels, corresponding to ALMA Band~5, the RF
  band 158--211\,GHz, and Band~9, the RF band 600--722\,GHz, provide
  state-of-the-art dual polarization receivers. The Band~5 frequency channel
  uses 2SB SIS mixers with an average SSB noise temperature around 45\,K
  with IF (intermediate frequency) band 4--8\,GHz for each sideband
  providing total $4\times 4$\,GHz IF band. The Band~9 frequency channel uses DSB SIS
  mixers with a noise temperature of 75--125\,K with IF band 4--12\,GHz for each
  polarization.}
{Both current SEPIA receiver channels are available to all APEX observers.}
{}


\maketitle

\section{Introduction}

APEX, the Atacama Pathfinder Experiment telescope, is a collaboration
between the Onsala Space Observatory (OSO), the European Southern
Observatory (ESO), and the Max-Planck-Institut für Radioastronomie
(MPIfR) to construct and operate a modified ALMA prototype antenna as
a single dish on the high altitude site (5105\,m) of the Llano
Chajnantor, Chile. The telescope was manufactured by VERTEX
Antennentechnik in Duisburg, Germany and is a 12\,m diameter
Cassegrain antenna with a surface r.m.s. better than
\SI{15}{\micro\metre}, two Nasmyth and a central Cassegrain cabins for
instrumentation. Details on the telescope, site, and instruments can
be found at \url{http://www.apex-telescope.org/} and in
\cite{gusten2006}. The new SEPIA (Swedish-ESO PI Instrument for the
APEX telescope) receiver was brought to APEX in the beginning of 2015
via a joint effort from GARD OSO and ESO. GARD performed the optics
and cryostat design and construction and the refurbishing of the
Band~5 pre-production cartridge (owned by ESO, earlier built by GARD
under European Community FP6 funded project) with the full-production LO system and
Warm Cartridge Assembly, both provided by NRAO via ESO. GARD has built
the control system and software, installed the SEPIA receiver in the
APEX Cabin A and performed technical commissioning. Initially, the
SEPIA receiver contained only an updated ALMA Band~5 pre-production
receiver cartridge \citep{bil2012} to address the growing interest in
future observations with Band~5 at ALMA (first available during cycle
5). The ALMA Band~5 center frequency nearly coincides with the
183.3\,GHz water absorption line. The key applications of the Band~5
receivers for ALMA range from observations of the 3$_{13}-2_{20}$
H$_2$O line at 183.310\,GHz in both Galactic objects and nearby
galaxies \citep[e.g.][]{hum2017} to the \SI{158}{\micro\metre}
emission line of C$^+$ from high redshift galaxies
\citep{laing2010}. In the beginning of 2016, SEPIA was equipped with
an additional ALMA Band~9 cartridge \citep{bar2015}, providing
observers with two state-of-the-art, single-pixel, dual polarization
receivers delivering each a total 16\,GHz IF band. The realization of
the SEPIA receiver project took about one year from the moment the
decision was taken, February-March 2014, to the time SEPIA was
installed at APEX in February 2015. The success of the SEPIA project
was guaranteed by ESO, providing optical windows and filters, access
to hardware from ALMA, support from NRAO by providing local oscillator
(LO), Band~5 warm cartridge assembly (WCA) and FE bias electronics --
all purchased by ESO. GARD OSO expertise in designing and constructing
mm-wave receivers, in particular the ALMA Band~5 receiver and a long
time heritage of APEX instrumentation, allowed the successful
completion of the entire project over this short time. In this
manuscript, the SEPIA receiver design, the receiver channel
performance and the commissioning results are described.

\section{SEPIA Receiver design}
\subsection{Optics}
\begin{figure}
  \begin{center}
    \includegraphics[width=\hsize]{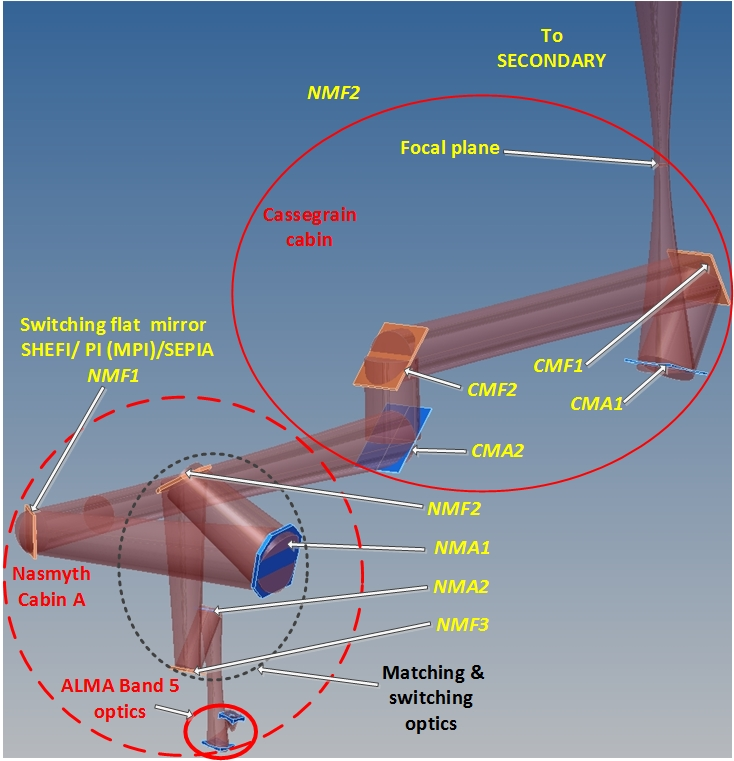}
  \end{center}

  \caption{SEPIA optics (in scale). Mirrors CMF1, CMA1, CMF2, CMA2 and
    Cabin A Instrument switch, NMF1, are part of earlier existing
    tertiary optics for the APEX multi-channel facility instrument
    SHeFI and PI instruments of the APEX Cabin A.  Mirrors NMA1, NMF2,
    NMF3 (SEPIA channel switch), NMA2 are parts of the SEPIA dedicated
    relay optics. Letters A and F in the mirror name indicate active
    and flat mirrors, respectively. Gaussian beam visualization is
    done for the lowest ALMA Band~5 RF frequency 158\,GHz. The entire
    relay optics (APEX Cabin A+C+SEPIA) provides frequency independent
    illumination of the secondary with $-12.5$\,dB edge taper and was
    designed to provide the beam clearance and optical element rims at
    $5\omega$.}
  \label{fig:optics}
\end{figure}

One of the major challenges of bringing the SEPIA receiver with
installed ALMA cartridges to the APEX telescope is the necessity to
implement tertiary optics, which should not only provide the required
and frequency independent illumination of the secondary but also be
compatible with the existing optical layout of the APEX Cabin A where
all single-pixel heterodyne receivers are installed. Another serious
constraint is a clearance of the APEX Cabin A Nasmyth tube, whose rim
is limited by the elevation encoder down to 150\,mm in diameter
requiring precision alignment possibilities to avoid Band~5 receiver
beam truncation. ALMA receiver cartridges have built-in cold optics
optimized for their respective position inside the ALMA Front-End (FE)
receiver cryostat place at the antenna focal plane. In particular,
depending on the cartridge position offset from the FE center, the
beam tilt offset compensating is different for, e.g., ALMA Band~5 --
2.38 degree and for ALMA Band~9 cartridges -- 0.93 degree. SEPIA
tertiary optics shall accommodate all these constrains and differences
with minimum number of reflecting surfaces (thus minimizing the
reflecting loss) and fit a very confined volume within the APEX Cabin
A. Specific for SEPIA and in contrast to ALMA FE, we use a rotating
cartridge-selection mirror. Such an optical switch addresses
limitations of the Nasmyth layout when one receiver channel has access
to the sky at a time (\cref{fig:optics}). The cartridge-selection
mirror (NMF3, \cref{fig:optics}) with its precision
computer-controlled rotating mechanism facilitates the accommodation
of different ALMA cartridges having specific differences in the
incoming beam positioning as outlined above.

The SEPIA optics has been designed and manufactured with its
supporting parts by GARD. The optics is mounted on a separate
supporting frame and employs precision mechanical reference devices;
these devices were used for initial alignment of the optics with help
of a reference laser. Once the optics as a whole is aligned with the
tertiary optics of the APEX Cabin A, the reference devices have been
locked to allow easier re-installation of the receiver optics and the
cryostat without further alignment. The SEPIA tertiary optics unit is
normally attached to the APEX Cabin A receiver supporting frame
whereas the SEPIA cryostat is attached to the optics-supporting
frame. The cryostat-optics frame mechanical interface allows accurate
re-mating of these two parts of the SEPIA receiver without disturbing
the optical alignment.

\subsection{SEPIA Cryostat}

\begin{figure}
  \begin{center}
    \includegraphics[width=\hsize]{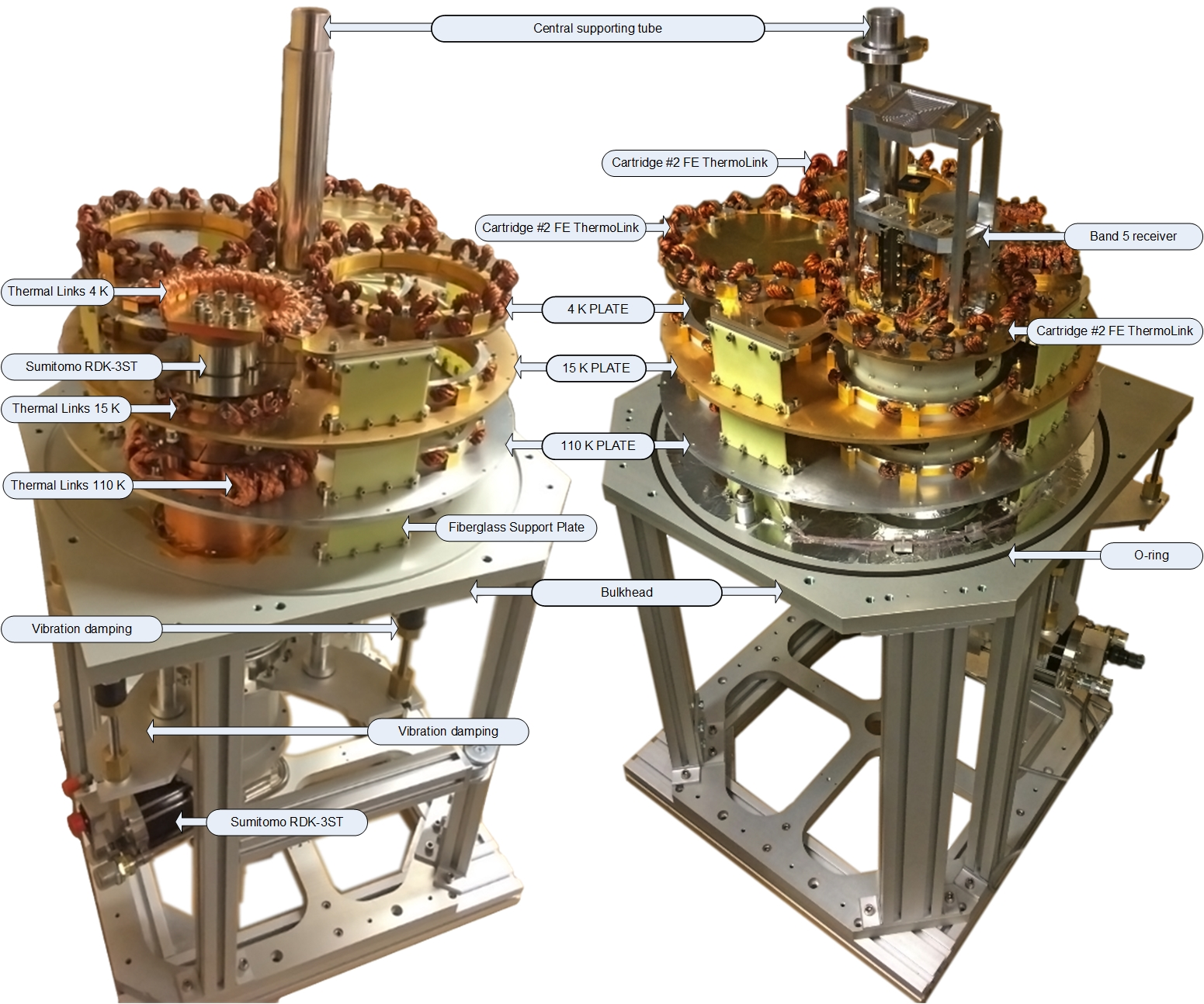}
  \end{center}
  \caption{SEPIA cryostat interior. Left: The cartridge cooling
    thermal links at 110K, 15K and 4K temperature stages (purchased
    from RAL, UK) and the Sumitomo cryo-cooler integrated into the
    cryostat. All thermal connections are made via oxygen-free copper
    flexible braids to avoid vibration transfer. Below the bulkhead,
    the elements of the anti-vibration cold-head suspension system are
    shown. Right: SEPIA cryostat interior with ALMA Band~5
    pre-production cartridge installed and the idle cartridge places
    blanked. Fiberglass support plates that also limit thermal
    conductivity flux sufficiently well providing the mechanical
    stability of the thermal decks.}
    \label{fig:cryo}
\end{figure}

In order to accommodate the ALMA receiver cartridges at APEX, a
compatible cryostat should be used which provides the required
mechanical, cryogenic, vacuum, optical and electrical interfaces to
the receiver cartridges. The space available in APEX Cabin A and the
cabin door opening completely excluded the use of the ALMA FE
cryostat. The Nasmyth layout of APEX allows using the receiver
cryostat in a steady position with the cryo-cooler and the receiver
cartridges placed vertically, thus achieving a more compact cryostat
design. A similar approach though with differently designed receiver
cartridges was used for the ASTE telescope \citep{sug2003}. The SEPIA
cryostat was built around a 3-stage Sumitomo RDK-3ST cryo-cooler
generation 6, with 1\,W at 4\,K lift off power. In order to provide
mechanical and cryogenic compatibility with the ALMA receiver
cartridges, we used the same distances and dimensions of the cartridge
holders as in the ALMA FE cryostat. Furthermore, the cartridge cooling
thermal links at 110K, 15K and 4K temperature stages have been
produced by Rutherford Appleton Laboratory, UK to the ALMA FE
specifications and used in the SEPIA cryostat. \Cref{fig:cryo},
left shows the interior of the SEPIA cryostat with the mounted
cartridge cooling thermal links and the Sumitomo cryo-cooler
integrated into the cryostat. At the center of the cryostat, we used a
stainless steel tube (protected by coaxial thermal shields) that
connects the bottom and top bulkheads of the cryostat vacuum vessel
and precludes the deflection of the bulkheads made off anodized
aluminum.

The cryostat vacuum vessel is made of two aluminum bulkheads and a
stainless steel tube interconnected using standard ISO-K claw clams.
The top bulkhead has three ports to connect the vacuum pump, vacuum
gauge and the venting valve that are integrated with the SEPIA
cryostat. All vacuum sealing is made with Viton\textsuperscript{TM}
O-rings. The SEPIA cryostat employs anti-vibration suspension of the
cryo-cooler by using a bellow between the 300K flange of the RDK-3ST
cryo-cooler and the cryostat to keep the vacuum with the
vibration-damping rubber elements installed at the brackets providing
the mechanical attachment of the cryo-cooler to the cryostat bottom
bulkhead (\Cref{fig:cryo}, left). All thermal links between the
cryo-cooler temperature stages and the elements of the receiver are
made with flexible thermal contacts using braided 0.05\,mm diameter
oxygen-free copper wires with each thermal link having about 6\,mm$^2$
cross-section area. This ensures that the vibration of the cryo-cooler
is not transferred to the receiver, its supporting structure and the
cryostat. The entire inner mechanical structure was simulated using
Ansys\textsuperscript{TM} FESS to verify that eigenfrequencies exceed
80\,Hz.

\begin{figure}
  \begin{center}
    \includegraphics[width=\hsize]{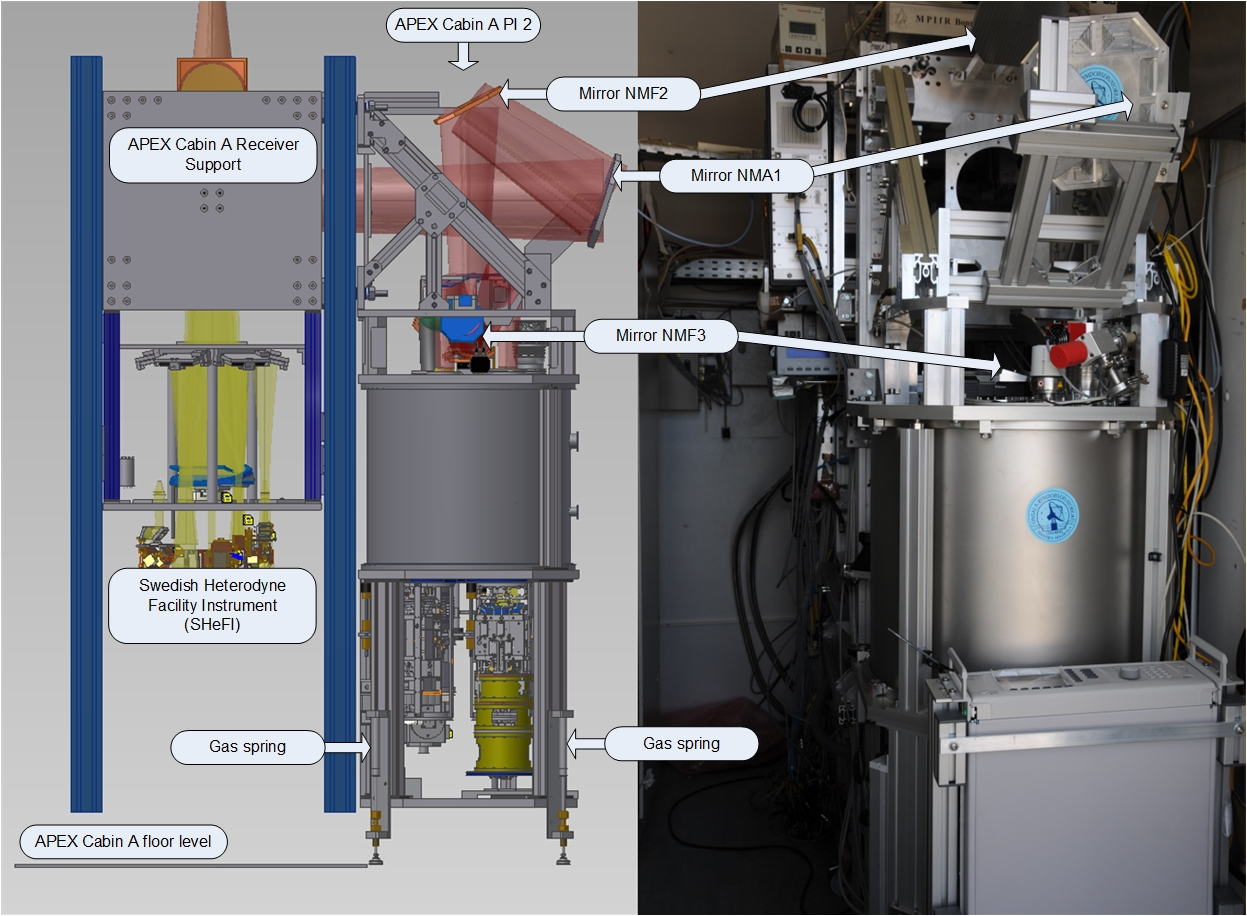}
  \end{center}

\caption{SEPIA receiver. Left: CAD picture showing the entire assembly
  of the SEPIA cryostat and dedicated SEPIA tertiary optics attached
  to the APEX Cabin A instrument-supporting frame. The SEPIA support
  towards the Cabin A floor uses gas springs unloading the frame and
  providing stable mechanical contact with the cabin floor. Right:
  picture of the SEPIA receiver installed at its outer PI~2 position
  in the APEX Cabin A.}
    \label{fig:rx}

\end{figure}

The total weight of the SEPIA receiver with tertiary optics and its
supporting frame, integrated turbo-pump and with three receiver
cartridges loaded is expected to be approximately 315\,kg. This also
includes three cartridge loader/lifter mounted at the receiver
cartridge loading interface that could be flexibly arranged to fit any
ALMA receiver cartridge. In order not to overload the APEX Cabin A
receiver support structure, a weight-compensation scheme was
implemented, comprising of the three supporting legs each equipped
with the rubber foot and joint to the SEPIA cryostat via
gas-spring. The gas springs each provide force at the level of $620\pm
20$\,N giving a total of 1860\,N of the weight compensation.
\Cref{fig:rx} shows a picture of the SEPIA receiver installed at
its PI~2 position in the APEX Cabin A.

The cryostat temperature is monitored and controlled via a
CryoCon\textsuperscript{TM} cryogenic temperature controller, Model
24C, which allows stabilizing the 4\,K temperature deck to facilitate
the SIS mixers gain stability and ensure that the sideband rejection
performance of the SIS mixers is not affected by changing the physical
temperature.  Additionally, the CryoCon\textsuperscript{TM} signal
from the cryogenic temperature sensors is used by the receiver control
system to interlock the receiver shut down at the events of a power
blackout, cold head or compressor failure when the raised temperature
does not allow operating SIS mixers.
A typical SEPIA cryostat cool-down time is around 8 hours reaching a
physical temperature of about 3.9 K\, in the laboratory conditions and
about 3.4\,K at the telescope, respectively. The cryo-refrigerator
capacity and SEPIA cryostat design allow operation of 3 cartridges
without substantial changes in the physical temperature.

The SEPIA cryostat RF vacuum windows have the same design as in the
ALMA FE cryostat, they are made of crystal quartz with Teflon coating
and were produced by QMC Inc. (for SEPIA cartridge Band~5 and
Band~9). The SEPIA cryostat uses the same infrared (IR) filters as the
corresponding bands of ALMA FE and those were delivered by ESO. The
design of the SEPIA RF vacuum window holders and the IR filter
brackets is such that it allows their exchange using a specialized
tool without disassembling the cryostat, thus providing full
flexibility of installing different frequency ALMA cartridges, i.e.,
any of the ALMA Band~5--10 cartridges, on three positions of the SEPIA
cryostat. The first channel pick-up mirror with its corresponding
support bracket should be individually introduced for each of the
(new) ALMA cartridges in order to accommodate them to the SEPIA
tertiary optics.

\subsection{SEPIA Receiver Control and Back End}

The SEPIA receiver control system hardware uses standard ALMA FE bias
modules (FEBM) for each receiver cartridge, which communicate with the
ALMA Monitoring and Control (M\&C) unit, all the hardware with its
respective firmware was developed and built by NRAO. The M\&C unit
communicates with the PC computer that is running the SEPIA control
software via CANbus using a PCAN-USB interface from PEAK. All
peripheral hardware communication with the control software is based
on an Ethernet internal network and communication with the APEX
Control System \citep[APECS, ][]{APEX-MPI-MAN-0011} is via a separate Ethernet
connection.
\Cref{fig:blocks} in the appendix shows the block diagram of the
SEPIA receiver control system, which consists of the receiver itself
together with all peripheral hardware.

The SEPIA control system PC is a dual boot system. Engineering
software running under MS Windows\textsuperscript{TM} 7 is based on
IRTECON \citep{erm2001} and used for advanced tuning and diagnostic of
the ALMA Band~5 receiver channel. For the ALMA Band~9 receiver
channel, NOVA uses specialized scripting software to tune the receiver
and check the hardware.

The observing control software is run under Linux and provided for all
SEPIA channels by OSO.  Tuning of the receivers is fully automatic and
controlled by APECS via network communication based on a command
language which is modelled after SCPI (Standard Commands for
Programmable Instruments). In addition, the program offers a graphical
user interface (GUI), allowing the observer to visually check the
state of the frontend, inspect entries in the tuning table and perform
manual tuning for testing purposes. The GUI also provides a plot of
the typical atmospheric transmission at Chajnantor over the frequency
band of the receiver, and marks current LO tuning and sideband
locations. Furthermore, a log of all SCPI commands as well as a
tabular view of raw CANbus readings are provided for debugging
purposes.

At present, the band 5 and band 9 receivers are controlled via two
independent programs, which share most of their code base, but differ
in parameters and hardware addressing.

The SEPIA receiver uses the XFFTS spectrometer \citep{kle2012}
provided as part of the APEX collaboration by MPIfR. The spectrometer
has 4--8\,GHz $\times$ 4 bandwidth, which covers 100\% of the SEPIA
ALMA Band~5 receiver channel IF band and only 4--8\,GHz $\times$ 2 of
the IF band of the SEPIA ALMA Band~9 receiver channel whereas the IF
band 8--12\,GHz is not available in the current configuration.
An upgrade of the IF system is planned for late 2018, which will cover 
4--12\,GHz IF bandwidth.

\subsection{SEPIA ALMA Band~5 Receiver Channel}

ALMA Band~5 receivers have been developed, designed and six
pre-production cartridges have been built under EC FP 6
``Infrastructure Enhancement Program''\footnote{ALMA Band 5, EC
  Framework Program 6 (FP6), under infrastructure enhancement,
  Contract 515906}. The details of the receiver cartridge design and
performance can be found in \citet{bil2012}. The produced
pre-production ALMA Band~5 cartridges have been delivered to ALMA via
ESO. When ALMA Band~5 full production started, it was decided to spare
the pre-production cartridges and one of those was made available by
ESO for the APEX SEPIA receiver. Several important modifications have
been made by GARD to the pre-production ALMA Band~5 cartridge to bring
it up-to-date. The most serious upgrade was done to the LO and WCA
\citep{bry2013}, the hardware version produced by NRAO for full ALMA
Band~5 production was used with appropriate changes in the cartridge
design.

\begin{figure}
  \begin{center}
    \includegraphics[width=\hsize]{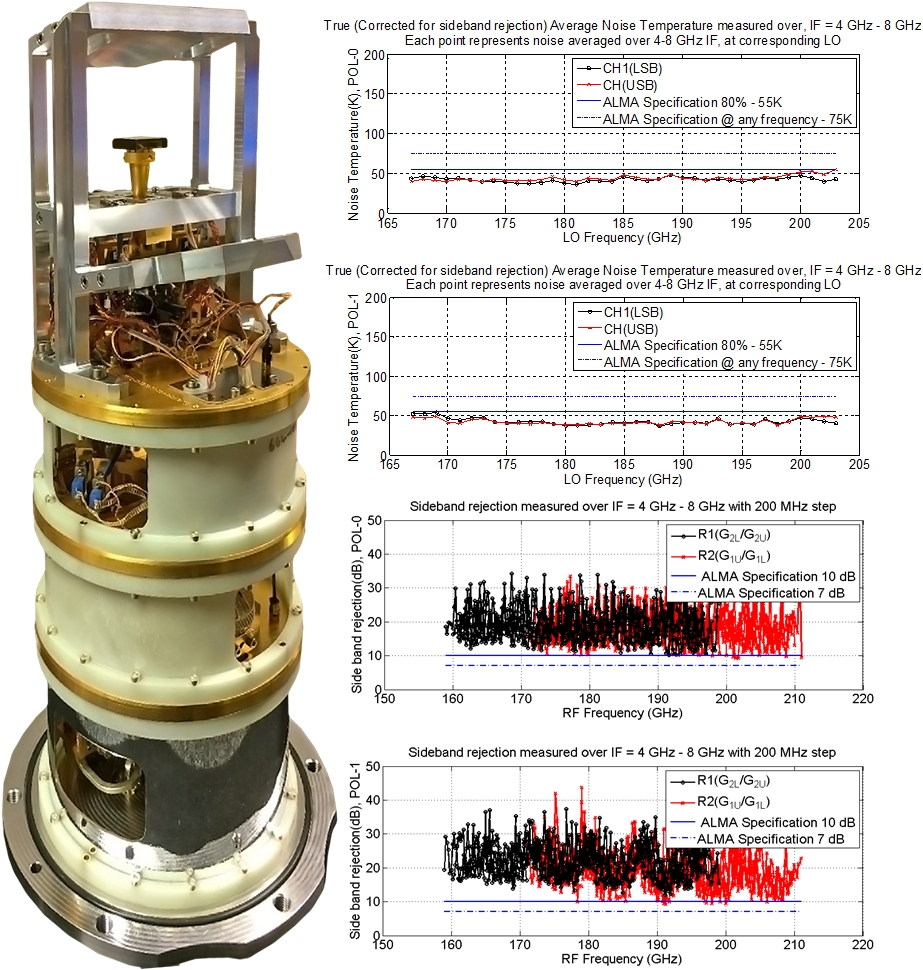}
  \end{center}

\caption{Left: Picture of the SEPIA ALMA Band~5 pre-production
  receiver cartridge. Right: Measured noise temperatures and the
  sideband rejection ratio of the SEPIA Band 5 receiver cartridge.}
    \label{fig:band5}
\end{figure}

The cartridge 2SB SIS mixer assemblies for both polarizations have
been replaced by the latest version with modifications of the 2SB SIS
mixers and IF hybrid assembly used for the full ALMA Band~5 production
project, e.g., a slightly extended RF band down to 158\,GHz that
allows to avoid the ``wings'' of the 183\,GHz water line when
observing with the upper sideband \citep{bel2017}. In
\Cref{fig:band5}, the picture of the modified ALMA Band~5
pre-production cartridge is shown along with the noise temperature and
sideband rejection ratio performance for both polarizations performed
in the lab keeping the SIS mixers at 4K physical temperature. As
mentioned above, the SEPIA cryostat is equipped with an active
temperature stabilization system, which allows stabilizing the SIS
mixer physical temperature down to milli-Kelvin level. This ensures
that the optimum 2SB mixer tuning setting for the best noise
temperature and the sideband-rejection will remain stable with changes
of the cold-head physical temperature which may occur, e.g., between
the cold head service periods.

Technical commissioning of the SEPIA receiver with the ALMA Band~5
receiver band installed was performed in February-March 2015. The
commissioning included hardware installation, e.g., reference LO PLL
synthesizer, IF switches, vacuum control gauge, turbo-pump,
computerized control system, etc., putting in place and alignment of
the SEPIA tertiary optics and installing the SEPIA cryostat and the
cold head compressor. The entire system was tested for integrity and
the SEPIA Band~5 beam alignment was verified using small size
Nitrogen-cooled absorbers at all tertiary optics components and at the
APEX Cabin C at the antenna focal plane.

\begin{table*}
  \caption{Summary of the SEPIA ALMA Band 5 specifications and performance.}

  \begin{tabular}{lll}
\hline
\hline
Item                     & Characteristics/Specs             & Comment                         \\
\hline
Technology               & 2SB SIS mixers, dual polarization &                                 \\
RF band                  & 157.36--211.64\,GHz               &                                 \\
IF band                  & 4--8\,GHz                         & For each sideband               \\
IF bandwidth             & 4 $\times$ 4\,GHz                 &                                 \\
SSB noise temperature    & $<$ 50\,K                         & Averaged over RF band $<$ 45\,K \\
Sideband rejection ratio & $>$ 10\,dB                        & Averaged over IF $>$ 16\,dB     \\
Cross-polarization       & $< -$23\,dB                       & At the cartridge RF window      \\
\hline
  \end{tabular}

\end{table*}

\subsection{SEPIA ALMA Band~9 Receiver Channel}

\begin{figure}
  \begin{center}
    \includegraphics[width=\hsize]{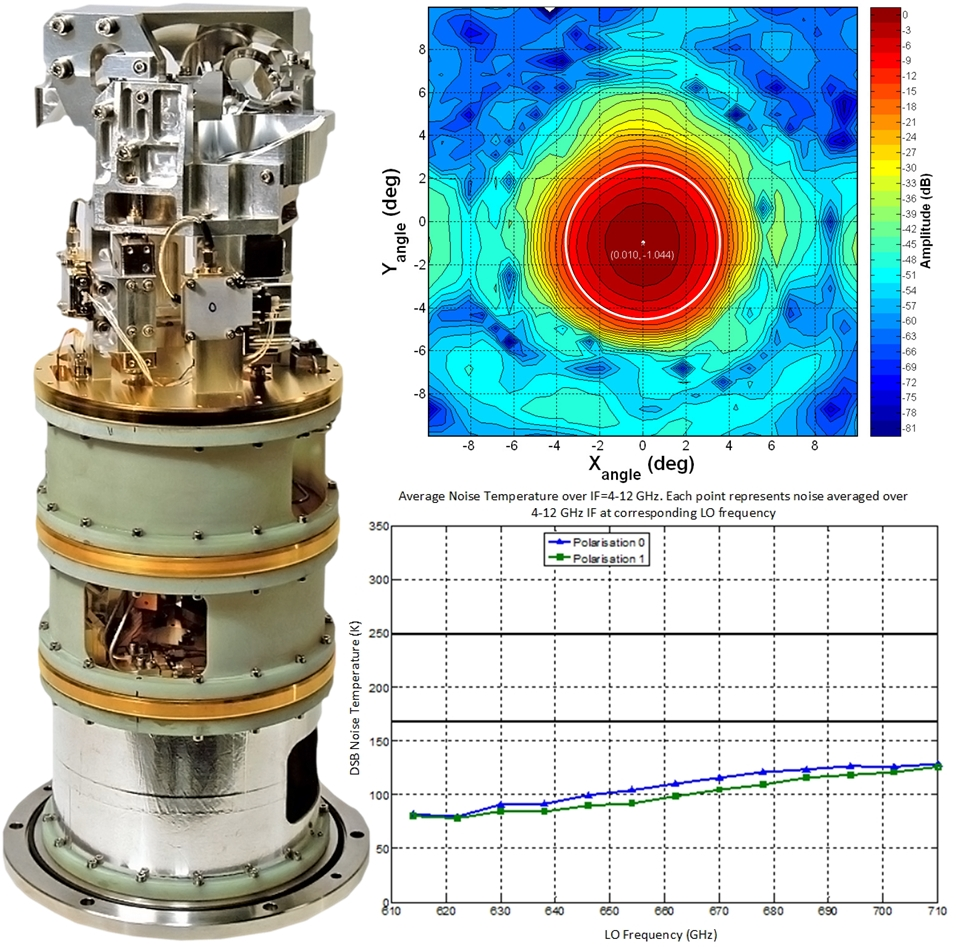}
  \end{center}

  \caption{Left: Picture of the SEPIA ALMA Band 9 pre-production
    receiver cartridge. Top right: Typical co-polar beam pattern in a
    Band 9 cartridge (polarization 0 in cartridge \#63, picture from
    \citet{bar2015}, with the secondary telescope mirror indicated
    (white circle). Bottom right: Noise temperatures of the SEPIA Band
    9 cartridge.}
  \label{fig:band9}

\end{figure}

The NOVA instrumentation group at the Kapteyn Astronomical Institute
in Groningen, The Netherlands, already produced a full set of ALMA
band 9 cartridges. For SEPIA, the NOVA group built another Band~9
cartridge at ALMA specifications and delivered it to the SEPIA PIs,
ESO and OSO. This SEPIA ALMA Band~9 receiver has DSB dual polarization
SIS mixers operating in the RF band of 599.77--722.15\,GHz. The
details on technology and performance of the ALMA Band~9 cartridge can
be found in \citet{bar2015}. The ALMA Band~9 LO, WCA and FE bias
module were produced by NRAO and jointly purchased by ESO and OSO for
the SEPIA project. \Cref{fig:band9} shows the picture of the
ALMA Band~9 production cartridge and the noise temperature for both
polarizations with the measurements performed in the lab keeping the
SIS mixers at 4K physical temperature for the ALMA Band~9 Cartridge
\#74 installed in SEPIA.

Technical commissioning of the SEPIA ALMA Band~9 receiver channel was
performed in February--March 2016 by a joint NOVA-GARD team. The
commissioning included hardware installation, IF switches,
computerized control system extension and software update. The entire
system was tested for integrity and the SEPIA Band~9 beam alignment
was verified using small size Nitrogen-cooled absorbers at all relay
mirrors and finally at the APEX Cabin C at the antenna focal plane.

The current plan is to replace the SEPIA ALMA Band~9 DSB receiver
channel in February-March 2018 with a new receiver cartridge, which
employs 2SB SIS mixers with four IF outputs, i.e. 4--12\,GHz USB and
LSB for each polarization, and otherwise has the same hardware and
layout as the ALMA Band~9 cartridge.

\begin{table*}
  \caption{Summary of the SEPIA ALMA Band 9 specifications and performance.}

  \begin{tabular}{lll}
\hline
\hline
Item                  & Characteristics/Specs             & Comment                                            \\
\hline
Technology            & DSB SIS mixers, dual polarization &                                                    \\
RF band               & 599.77--722.15\,GHz               &                                                    \\
IF band               & 4--12\,GHz                        & DSB operation                                      \\
IF bandwidth          & 2 $\times$ 8\,GHz                 & Currently limited by the FFTS to 2 $\times$ 4\,GHz \\
DSB noise temperature & 75--125\,K                        & Averaged over RF band $\approx$100\,K              \\
Cross-polarization    & $-$15.5\,dB                       & At the cartridge RF window                         \\
\hline
  \end{tabular}

\end{table*}

\section{Sky commissioning}

\begin{figure}
  \begin{center}
    \includegraphics[width=\hsize]{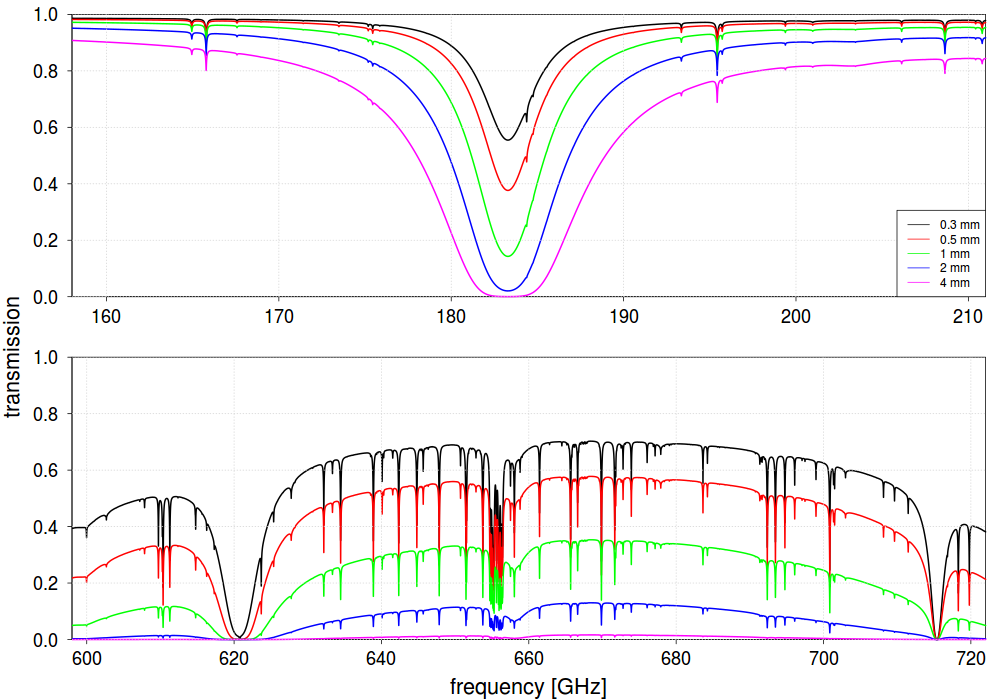}
  \end{center}
  \caption{Atmospheric transmission at Llano de Chajnantor
    \citep{paine_scott_2017_438726} for the frequency ranges of the
    band 5 (upper panel) and band 9 (lower panel) receivers.}
  \label{Atmosphere}
\end{figure}

The sky commissioning of the SEPIA Band 5 receiver was successfully
performed between February and September 2015. The sky commissioning
of Band 9 has been started in June 2016.

The atmospheric transmission at Llano de Chajnantor in the frequency
range of the SEPIA Band 5 and Band 9 receivers is shown in
\cref{Atmosphere}.

A number of on-sky tests have to be conducted before a new instrument
can be offered for observing time. Tuning Band 5 every 0.5\,GHz, we
confirmed that the whole tuning range from 157.36--211.64\,GHz is
accessible. Observing strong water masers as part of the science
verification, \citep{hum2017} determined the sideband suppression
level to 17.7 dB from comparing the water line in the upper sideband
to its ghost line in the lower sideband.

To determine the beam coupling efficiencies for Band 5, we observed
Mars on UT 2016 April 6 and 7 at 208\,GHz, and Jupiter on UT 2016
August 11 at 170\,GHz.  The respective apparent sizes of these planets
were 12.7$^{\prime \prime}$ for Mars and 31.4$^{\prime \prime}$ for
Jupiter.  From the same planet observations, we derive deconvolved
beam sizes of about 28.6$^{\prime \prime}$ at 208\,GHz and
31.7$^{\prime \prime}$ at 170\,GHz.  This means the apparent size of
Jupiter matched the beam size, while Mars was an unresolved source
(semi-extended in the {\it Herschel}/HIFI nomenclature).  Assuming an
antenna far-field forward efficiency $\eta_{f}$=0.95 (from
\citealt{gusten2006}, with 10\% uncertainties across the 160-211 GHz
range), we find beam coupling efficiencies of 0.67$\pm$0.05 for
Jupiter and 0.72$\pm$0.07 for Mars, corresponding to Jy/K factors of
33.8$\pm$2.5 and 38.4$\pm$2.8 for resolved and unresolved sources,
respectively.

To determine the beam coupling efficiencies for Band 9, we observed
Jupiter on UT 2017 April 22, and Uranus on UT 2017 May 21, both at
691.5\,GHz.  The apparent sizes for Uranus and Jupiter were
3.4$^{\prime \prime}$ and 42.6$^{\prime \prime}$, respectively. From
observations toward Callisto, which has an apparent size of
1.5$^{\prime \prime}$, we determine a deconvolved beam size of
8.8$\pm$0.5$^{\prime \prime}$ at 691.5\,GHz.  Assuming the same
antenna far-field forward efficiency $\eta_{f}$=0.95 as above, we find
beam coupling efficiencies of 0.46$\pm$0.02 for Jupiter and
0.37$\pm$0.03 for Uranus, corresponding to Jy/K factors of 63$\pm$3
and 79$\pm$6, for resolved and unresolved sources, respectively.


These values were obtained after careful optimization of the antenna
performance (replacement of the secondary surface and dish setting
following holography results).  At these high frequencies, the
effective performance depends critically on both the antenna and the
instrument.  While we expect the performance of the instrument to be
relatively stable, the antenna efficiencies can vary on time scales of
weeks or months, due to the external weather factors.  Users are
encouraged to perform efficiency measurements as part of their science
projects, or to contact APEX staff for the most recent values to be
used.

\begin{figure}
  \begin{center}
    \includegraphics[width=\hsize]{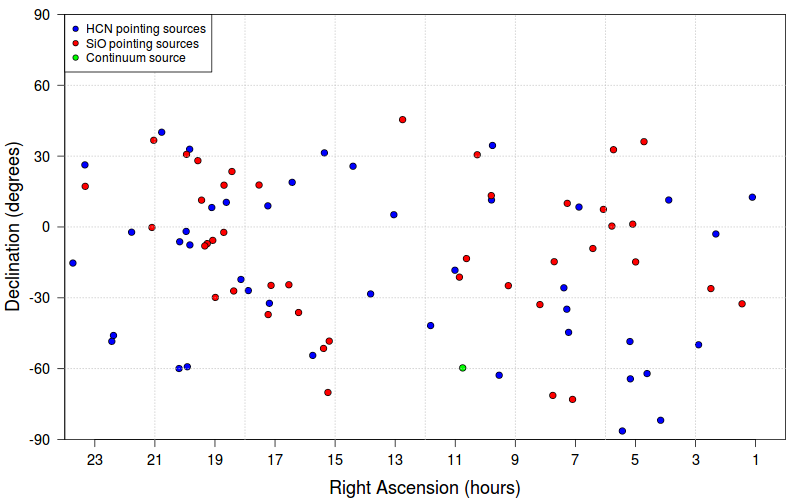}
  \end{center}
  \caption{Sky distribution of SEPIA Band 5 pointing sources (see
    \cref{Pointing-Sources}).}
  \label{fig:pointing}
\end{figure}

Pointing and focus observations at APEX are conducted towards compact
sources with either strong continuum or line emission. For the other
APEX heterodyne instruments, these observations are often done in the
main transitions of the CO molecule. Since the Band 5 frequency range
does not contain a CO transition, we had to find sources that strongly
emit in other molecular transitions. We chose evolved stars that are
bright in HCN\,(2--1) or SiO\,(4--3){\it v}[0--3] masers. After
testing their quality as pointing sources, we constructed a pointing
catalog, which covers the whole LST range (\cref{fig:pointing},
\cref{Pointing-Sources}). The pointing sources are being used to
compile a good pointing model for Band 5, using the professional
version of the Telescope Pointing Analysis System
TPOINT\footnote{http://www.tpointsw.uk} \citep{wallace1994}. With this
receiver, pointing accuracies better than 2.5" are achieved.

\begin{figure}
  \begin{center}
    \includegraphics[width=\hsize]{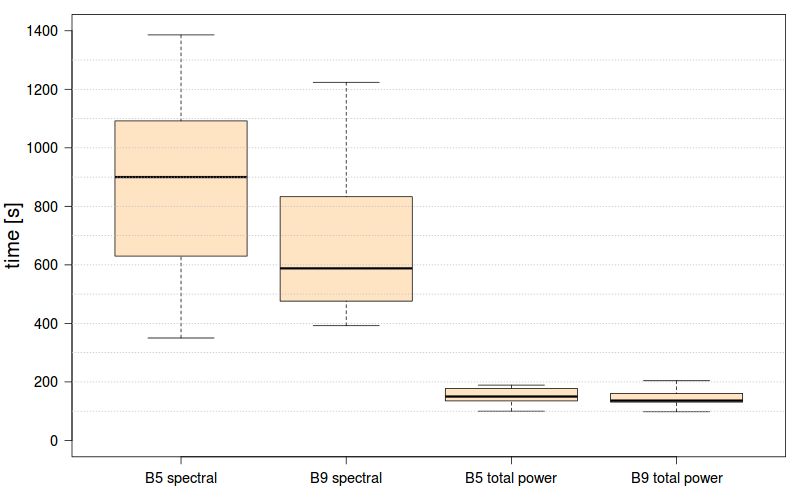}
  \end{center}
  \caption{Results from Allan variance measurements. The bold
    horizontal lines mark the median results of all measurements
    taken, the top and bottom of the boxes show the 25th and 75th
    percentile, i.e. outline the middle 50\% of the data.}
  \label{fig:stability}
\end{figure}

The receiver stability was tested using a series of Allan variance
measurements, both in spectral and total power mode. Results are shown
in \cref{fig:stability}. For all measurements, a channel
resolution of 1.2\,MHz was used. For both Band 5 and Band 9
spectroscopic stability times are always well above a few times 100
seconds. For the quoted resolution, even the total power stability is
typically above 100 seconds, or above 30 seconds when scaled to
10\,MHz resolution.

\section{Calibration of Band 5 data}

The online calibration of spectral line observations at the APEX
telescope is based on SKY-HOT-COLD calibration scans immediately
before the spectral line data is taken. During these scans,
observations of the blank sky, a hot load at ambient temperature and a
cold load at liquid nitrogen temperature are conducted. From the HOT
and COLD signal, the Online Calibrator determines the receiver
temperature. From the SKY phase, it calculates the sky temperature,
correcting for spillover and forward efficiency. Based on the
elevation of the SKY observations and a sophisticated atmospheric
model, ATM \citep{pardo2001}, the opacity in image and signal band is
determined. In the online calibration, an average opacity is
calculated based on 10 equally spaced frequencies across the 2.5\,GHz
band pass.  This is sufficient if the atmosphere is reasonably
flat. However, in Band 5 due to the water absorption of the
atmosphere, this approach is insufficient for observations close to
183.3\,GHz. This type of data needs to be recalibrated offline at the
telescope. In the offline calibration, opacity values are determined
in chunks of 128 channels (the resolution of the atmospheric
model). The online calibrated data underestimates the emission around
183\,GHz by up to 30\% \citep[e.g.][]{Immer2016}. The goal is to
upgrade the online calibration to provide such optimally calibrated data
straight away.

\section{Science with SEPIA Band 5}

After commissioning and science verification, over 100 regular
proposals have been accepted on SEPIA in the first two years of
operations. The SEPIA proposals represent between 25\% and 50\% of the
accepted proposals in the past four observing periods, and over 1000
hours have already been observed. Seven science papers have
been published at the time of writing, and a science meeting
describing these results was held at ESO \citep[]{debreuck17}.\footnote{see also
  \url{https://www.eso.org/sci/meetings/2017/band5/program.html}}

One of the main science drivers for an ALMA Band 5 receiver at APEX
was the possibility to observe the 183.3\,GHz water line. Interstellar
water emission was first detected by \citep{cheung69} and since then,
the Herschel Space Observatory has shown that gaseous water is
widespread in molecular clouds in our Galaxy \citep{vandishoeck11}.
With APEX, surveys for both maser as well as thermal water emission
towards a large number of sources are feasible.  The water molecule is
an important tracer of the energetic processes which take place during
the formation of low- and high-mass stars. In combination with the
lower optical depth H$_2^{18}$O transition at 203.4\,GHz, water
observations probe the water content even in the inner parts of hot
molecular cores, complementing space-based water observations
(e.g. from Herschel). The 183.3\,GHz line has also been detected in
the O-rich circumstellar envelopes of several evolved stars with
mass-loss rates exceeding 10$^{-6}$ M$_\odot$/yr \citep{gon1998,
  gon1999}. Observations of this transition with APEX will thus allow
the study of the water abundance in evolved stars. In addition, the
degree of the linear polarization of the 183.3\,GHz water maser
emission can be determined from dual polarization observations with
Band 5 \citep{hum2017}. Furthermore, water maser observations will
provide information about the water content of nearby galaxies ($z <
0.1$) as well as the high temperature and density regions of AGNs.  In
the Seyfert 2 galaxy NGC 4945, 183.3\,GHz H$_2$O megamaser emission
was detected for the first time with SEPIA Band 5 \citep{hum2016}.  It
is the strongest extragalactic submillimeter water maser detected up
to date. The emission seems to be dominated by that from the AGN
central engine.

\citet{gal2016} recently detected water (183.3\,GHz) and the methanol
($4-3$) group (193.5\,GHz) in the ultra-luminous infrared galaxy
Arp~220. No time variations are observed in the megamaser water line
compared to previous observations.  This supports arguments against an
AGN nuclear origin for the line. The J$=2-1$ transitions of the
nitrogen-bearing species HCN, HNC, and N$_2$H$^+$ as well as HCO$^+$
are available in the SEPIA Band 5 frequency range. Their isotopologues
H$^{13}$CN, HC$^{15}$N, H$^{13}$CO$^{+}$, and HC$^{18}$O$^{+}$ are
observable in one frequency setting as are HCN and HCO$^{+}$. These
transitions serve as low-energy complements to the higher J-lines
which are covered by other APEX bands, probing colder/lower density
material. In evolved stars, simultaneous observations of HCN and SiO
allow the determination of the HCN/SiO intensity ratio which is
indicative of the star's chemical type \citep{olo1998}.

The symmetric top molecules CH$_3$CN and CH$_3$CCH have several
K-ladders in the band. Also, the methanol J$=4-3$ line forest around
193.5\,GHz and the set of lines of CCH($2-1$) around 174.7\,GHz are
accessible with SEPIA Band 5. All these transitions are excellent
temperature tracers of the denser molecular gas. They are more
sensitive to lower temperatures than the corresponding transitions in
other APEX bands. \citet{molinari16} observed CH$_3$CCH ($12-11$)
towards 51 dense clumps and determined their gas temperatures. Many
other molecules have transitions in the SEPIA Band 5 frequency
range. \Cref{Transitions-Band5} lists some of the astrophysically
important ones.

For all sources at $z < 0.615$, at least one CO transition falls
within the frequency range of the SEPIA Band 5 receiver. This allows
observations of high-J CO transitions in bright submillimeter galaxies
to investigate their CO spectral line energy distributions or confirm
their redshifts. \citet{str2016}, who studied the redshift
distribution of gravitationally lensed dusty star-forming galaxies 
discovered from the South Pole Telescope, used the SEPIA CO observations of two
sources to confirm their redshifts. Detecting [C{\sc II}] in highly
redshifted sources ($z < 8.9$) is challenging for APEX but extremely
rewarding. 

Due to the 8\,GHz bandwidth per polarization channel, the
whole frequency range of SEPIA Band 5 can be covered with only eight
frequency tunings with several 100\,MHz overlap. As ESO science
verification project, Immer et al. (2017), in prep. observed the
frequency range 159.2--210.7\,GHz towards the high-mass star forming
complex Sgr~B2, proving once again the line richness of this source
and illustrating the large number of transitions that can be detected
in SEPIA Band 5.

\begin{table*}
  \caption{Astrophysically important molecular transitions within the
    frequency coverage of SEPIA Band 5.}
  \label{Transitions-Band5}
  \begin{tabular}{cccc}
Molecule         & Transition              & Frequency (GHz) & Comment              \\
CH$_{3}$CN       & J=9$-$8                           & 165.60 & Several K-components \\
H$_{2}$S         & J$_{Ka,Kc}$ = 1$_{1,1}-$1$_{0,1}$ & 168.76 &                      \\
HC$^{18}$O$^{+}$ & J=2$-$1                           & 170.32 &                      \\
CH$_{3}$CCH      & J=10$-$9                          & 170.90 & Several K-components \\
SiO, v=2         & J=4$-$3                           & 171.28 &                      \\
HC$^{15}$N       & J=2$-$1                           & 172.11 &                      \\
SiO, v=1         & J=4$-$3                           & 172.48 &                      \\
H$^{13}$CN       & J=2$-$1                           & 172.68 &                      \\
H$^{13}$CO$^{+}$ & J=2$-$1                           & 173.51 &                      \\
SiO, v=0         & J=4$-$3                           & 173.69 &                      \\
CCH              & N=2$-$1                           & 174.70 & Several lines        \\
HCN              & J=2$-$1                           & 177.26 &                      \\
HCO$^{+}$        & J=2$-$1                           & 178.38 &                      \\
HNC              & J=2$-$1                           & 181.32 &                      \\
H$_{2}$O         & J$_{Ka,Kc}$ = 3$_{1,3}-$2$_{2,0}$ & 183.31 &                      \\
CH$_{3}$CN       & J=10$-$9                          & 183.90 & Several K-components \\
$^{13}$CS        & J=4$-$3                           & 184.98 &                      \\
N$_{2}$H$^{+}$   & J=2$-$1                           & 186.34 &                      \\
CH$_{3}$CCH      & J=11$-$10                         & 187.90 & Several K-components \\
C$^{34}$S        & J=4$-$3                           & 192.82 &                      \\
CH$_{3}$OH       & J=4$-$3                           & 193.50 & Several lines        \\
CS               & J=4$-$3                           & 195.95 &                      \\
CH$_{3}$CN       & J=11$-$10                         & 202.30 & Several K-components \\
H$_{2}^{18}$O    & J$_{Ka,Kc}$ = 3$_{1,3}-$2$_{2,0}$ & 203.41 &                      \\
CH$_{3}$CCH      & J=12$-$11                         & 205.00 & Several K-components \\
  \end{tabular}
\end{table*}

\begin{table*}
  \caption{Astrophysically important molecular transitions within the
    frequency coverage of SEPIA Band 9.}
  \label{Transitions-Band9}
  \begin{tabular}{cccc}
Molecule         & Transition              & Frequency (GHz) & Comment              \\
H$^{13}$CO+      & J=7$-$6                           & 607.18 &                     \\
SiO              & J=14$-$13                         & 607.61 &                     \\
HN$^{13}$C       & J=7$-$6                           & 609.51 &                     \\
HCN-v2           & J=7$-$6                           & 623.36 &                     \\
HCO$^+$          & J=7$-$6                           & 624.21 &                     \\
SiH              & $^2\Pi F_1;J=3/2^+ $-$ J=1/2^+$   & 624.92 & Several lines       \\
H$^{37}$Cl       & J=1$-$0                           & 624.96 & Several lines       \\
H$^{35}$Cl       & J=1$-$0                           & 625.90 & Several lines       \\
C$^{34}$S        & J=13$-$12                         & 626.34 &                     \\
HNC              & J=7$-$6                           & 634.51 &                     \\
$^{30}$SiO       & J=15$-$14                         & 635.22 &                     \\
CS               & J=13$-$12                         & 636.53 &                     \\
$^{29}$SiO       & J=15$-$14                         & 642.80 &                     \\
SiO              & J=15$-$14                         & 650.96 &                     \\
DCN              & J=9$-$8                           & 651.57 &                     \\
C$^{18}$O        & J=6$-$5                           & 658.55 &                     \\
$^{13}$CO        & J=6$-$5                           & 661.07 &                     \\
C$^{17}$O        & J=6$-$5                           & 674.01 &                     \\
$^{30}$SiO       & J=16$-$15                         & 677.51 &                     \\
HCS$^+$          & J=16$-$15                         & 682.44 &                     \\
CS               & J=14$-$13                         & 685.43 &                     \\
$^{29}$SiO       & J=16$-$15                         & 685.59 &                     \\
HC$^{15}$N       & J=8$-$7                           & 688.28 &                     \\
H$^{13}$CN       & J=8$-$7                           & 690.55 &                     \\
CO               & J=6$-$5                           & 691.47 &                     \\
H$^{13}$CO$^+$   & J=8$-$7                           & 693.88 &                     \\
SiO              & J=16$-$15                         & 694.29 &                     \\
HN$^{13}$C       & J=8$-$7                           & 696.53 &                     \\
HCN-v2           & J=8$-$7                           & 708.79 & Several lines       \\
HCN              & J=8$-$7                           & 708.88 &                     \\
HCO$^+$          & J=8$-$7                           & 713.34 &                     \\
$^{30}$SiO       & J=17$-$16                         & 719.79 &                     \\
C$^{34}$S        & J=15$-$14                         & 722.58 &                     \\
\end{tabular}
\end{table*}

\section{Conclusion}

In this paper, we describe the APEX SEPIA receiver which was installed
in APEX Cabin~A at the beginning of 2015.  The SEPIA receiver
extensively uses ALMA technology developments and is a flexible
platform, which allows the use of up to three regular ALMA receiver
cartridges and new receivers.  Currently, SEPIA provides APEX
observers with two state-of-the-art dual-polarization receivers,
covering two frequency channels 158--211\,GHz, ALMA Band 5, and
600--720\,GHz, ALMA Band 9. The excellent performance of the receiver
was confirmed during science commissioning.  Besides its advanced
latest technology and current capabilities, SEPIA provides an
excellent opportunity of testing and verifying experimental receivers
for future ALMA upgrades, e.g., planned in 2018 an upgrade of SEPIA
ALMA Band 9 receiver channel with 2SB SIS mixers. We expect that
coming upgrades will further improve the receiver and give greater
opportunities for observing with APEX. All current SEPIA receiver
channels are available to all APEX observers, and all data are
publicly available after a 1 year proprietary period.

\begin{acknowledgements}
We would like to thank the APEX science team for valuable assistance
during on-sky commissioning of the new receiver.  Prof. K. Menten,
MPIfR, is acknowledged for making available the XFFTS spectrometer for
use with SEPIA. We would like to acknowledge Kamaljeet S. Saini and
NRAO for providing the LO and Band 5 and Band 9 warm cartridge
hardware. We also thank Karina Celed\'on and Alicia Garafulich for
their help with the logistics.
\end{acknowledgements}

\bibliographystyle{aa}
\bibliography{SEPIA}

\begin{appendix}

\setcounter{figure}{0}
\renewcommand\thefigure{\Alph{section}.\arabic{figure}}

\section{Block diagram}
\begin{figure}[H]
  \centering
  \includegraphics[width=\textwidth]{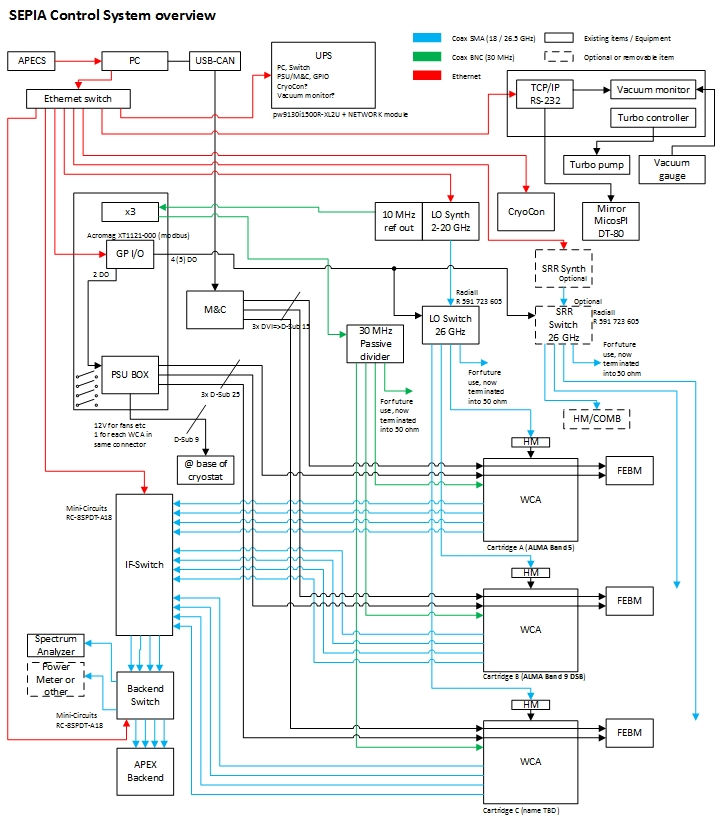}

  \caption{Block diagram of the SEPIA receiver control system with all
    peripheral hardware.}
  \label{fig:blocks}

\end{figure}

  \section{Pointing sources}
\longtab[1]{
\begin{longtable}{lr@{:}lr@{:}lcc}
  \caption{Coordinates of the SEPIA Band 5 pointing sources, obtained
    from different catalogs.}
  \label{Pointing-Sources}\\
Name & \multicolumn{2}{c}{RA [hms]} & \multicolumn{2}{c}{Dec [$^{\circ\prime\,\prime\prime}$]} & Molecule & Catalog\\
WX-Psc         & 01&06:25.98  & $+$12&35:53.0  & SiO & APEX SHeFI\\
R-Scl          & 01&26:58.079 & $-$32&32:35.76 & HCN & APEX SHeFI\\
o-Ceti         & 02&19:20.790 & $-$02&58:41.91 & SiO & APEX SHeFI\\
R-For          & 02&29:15.325 & $-$26&05:55.65 & HCN & APEX SHeFI\\
R-Hor          & 02&53:52.899 & $-$49&53:22.28 & SiO & APEX SHeFI\\
IK-Tau         & 03&53:28.89  & $+$11&24:21.8  & SiO & APEX SHeFI\\
U-Men          & 04&09:35.892 & $-$81&51:17.57 & SiO & SIMBAD, Optical\\
R-Dor          & 04&36:45.499 & $-$62&04:38.51 & SiO & APEX SHeFI\\
CRL618         & 04&42:53.64  & $+$36&06:53.4  & HCN & APEX SHeFI\\
R-Lep          & 04&59:36.353 & $-$14&48:22.53 & HCN & APEX SHeFI\\
W-Ori          & 05&05:23.723 & $+$01&10:39.46 & HCN & APEX SHeFI\\
U-Dor          & 05&10:08.845 & $-$64&19:04.27 & SiO & SIMBAD, Optical\\
S-Pic          & 05&10:57.248 & $-$48&30:25.45 & SiO & SIMBAD, Optical\\
R-Oct          & 05&26:06.196 & $-$86&23:17.77 & SiO & SIMBAD, Optical\\
V370-Aur       & 05&43:49.68  & $+$32&42:06.2  & HCN & APEX SHeFI\\
N2071IR        & 05&47:04.85  & $+$00&21:47.1  & HCN & APEX LABOCA\\
RAFGL865       & 06&04:00.0   & $+$07&25:53.0  & HCN & APEX SHeFI\\
V636-Mon       & 06&25:01.43  & $-$09&07:15.9  & HCN & APEX SHeFI\\
GX-Mon         & 06&52:47.05  & $+$08&25:18.7  & SiO & APEX SHeFI\\
R-Vol          & 07&05:36.20  & $-$73&00:52.0  & HCN & APEX SHeFI\\
L2-Pup         & 07&13:32.42  & $-$44&38:19.8  & SiO & APEX SHeFI\\
HD56126        & 07&16:10.260 & $+$09&59:47.93 & HCN & APEX SHeFI\\
SVS-03513      & 07&17:05.68  & $-$34&49:39.0  & SiO & SEST\\
VY-CMa         & 07&22:58.321 & $-$25&46:03.49 & SiO & APEX SHeFI\\
OH231.8        & 07&42:16.92  & $-$14&42:50.0  & HCN & APEX SHeFI\\
07454-7112     & 07&45:02.41  & $-$71&19:45.7  & HCN & APEX SHeFI\\
RAFGL1235      & 08&10:48.90  & $-$32&52:05.5  & HCN & APEX SHeFI\\
RAFGL5254      & 09&13:53.95  & $-$24&51:25.2  & HCN & APEX SHeFI\\
R-Car          & 09&32:14.601 & $-$62&47:19.97 & SiO & SIMBAD, Optical\\
R-LMi          & 09&45:34.293 & $+$34&30:42.75 & SiO & APEX SHeFI\\
R-Leo          & 09&47:33.494 & $+$11&25:43.21 & SiO & APEX SHeFI\\
IRC+10216      & 09&47:57.41  & $+$13&16:43.6  & HCN & APEX SHeFI\\
RW-LMi         & 10&16:02.29  & $+$30&34:19.1  & HCN & APEX SHeFI\\
U-Hya          & 10&37:33.300 & $-$13&23:04.74 & HCN & APEX SHeFI\\
eta Carina     & 10&45:03.53  & $-$59&41:03.75 & Continuum & APEX LABOCA\\
V-Hya          & 10&51:37.248 & $-$21&15:00.38 & HCN & APEX SHeFI\\
R-Crt          & 11&00:33.828 & $-$18&19:29.6  & SiO & APEX SHeFI\\
X-Cen          & 11&49:11.788 & $-$41&45:27.27 & SiO & SIMBAD, Optical\\
Y-Cvn          & 12&45:07.828 & $+$45&26:24.92 & HCN & SIMBAD, Optical\\
RT-Vir         & 13&02:38.007 & $+$05&11:08.24 & SiO & APEX SHeFI\\
W-Hya          & 13&49:01.961 & $-$28&22:04.13 & SiO & APEX SHeFI\\
RX-Boo         & 14&24:11.644 & $+$25&42:12.93 & SiO & APEX SHeFI\\
RAFGL4211      & 15&11:41.45  & $-$48&19:59.0  & HCN & APEX SHeFI\\
X-TrA          & 15&14:19.180 & $-$70&04:46.17 & HCN & APEX SHeFI\\
S-CrB          & 15&21:23.947 & $+$31&22:02.39 & SiO & APEX SHeFI\\
IRAS15194-5115 & 15&23:05.07  & $-$51&25:58.7  & HCN & APEX SHeFI\\
IRSV1540       & 15&44:39.803 & $-$54&23:05.03 & SiO & SIMBAD, Infrared\\
NGC6072        & 16&12:58.079 & $-$36&13:46.06 & HCN & APEX SHeFI\\
U-Her          & 16&25:47.472 & $+$18&53:32.86 & SiO & SIMBAD, Optical\\
IRAS16293      & 16&32:22.56  & $-$24&28:31.8  & HCN & SIMBAD, Infrared\\
RAFGL1922      & 17&07:58.11  & $-$24&44:31.2  & HCN & APEX SHeFI\\
AH-Sco         & 17&11:17.021 & $-$32&19:30.71 & SiO & SIMBAD, Optical \\
NGC6302        & 17&13:44.46  & $-$37&06:10.7  & HCN & APEX SHeFI\\
V2108-Oph      & 17&14:19.393 & $+$08&56:02.60 & SiO & SIMBAD, Infrared\\
IRC+20326      & 17&31:55.30  & $+$17&45:21.0  & HCN & APEX SHeFI\\
OH2.6-0.4      & 17&53:18.92  & $-$26&56:37.1  & SiO & SEST\\
VX-Sgr         & 18&08:04.048 & $-$22&13:26.63 & SiO & SIMBAD, Optical\\
RAFGL2135      & 18&22:34.68  & $-$27&06:29.4  & HCN & APEX SHeFI\\
RAFGL2155      & 18&26:05.84  & $+$23&28:46.7  & HCN & APEX SHeFI\\
V1111-Oph      & 18&37:19.26  & $+$10&25:42.2  & SiO & APEX SHeFI\\
IRC+20370      & 18&41:54.55  & $+$17&41:08.6  & HCN & APEX SHeFI\\
IRC+00365      & 18&42:24.87  & $-$02&17:27.2  & HCN & APEX SHeFI\\
IRC-30398      & 18&59:13.851 & $-$29&50:20.44 & HCN & SIMBAD, Optical\\
V-Aql          & 19&04:24.162 & $-$05&41:05.43 & HCN & APEX SHeFI\\
R-Aql          & 19&06:22.254 & $+$08&13:47.25 & SiO & APEX SHeFI\\
W-Aql          & 19&15:23.365 & $-$07&02:50.36 & HCN & APEX SHeFI\\
IRC-10502      & 19&20:18.12  & $-$08&02:12.0  & HCN & APEX SHeFI\\
V1302-Aql      & 19&26:48.03  & $+$11&21:16.7  & HCN & APEX SHeFI\\
V1965-Cyg      & 19&34:10.05  & $+$28&04:08.5  & HCN & APEX SHeFI\\
GY-Aql         & 19&50:06.329 & $-$07&36:52.52 & SiO & APEX SHeFI\\
chi-Cyg        & 19&50:33.905 & $+$32&54:50.19 & SiO & APEX SHeFI\\
S-Pav          & 19&55:13.967 & $-$59&11:44.34 & SiO & SIMBAD, Optical\\
RAFGL2477      & 19&56:48.45  & $+$30&44:02.6  & HCN & APEX SHeFI\\
RR-Aql         & 19&57:36.044 & $-$01&53:11.80 & SiO & APEX SHeFI\\
IRC-10529      & 20&10:27.87  & $-$06&16:13.6  & SiO & APEX SHeFI\\
X-Pav          & 20&11:45.896 & $-$59&56:12.95 & SiO & APEX SHeFI\\
NML-Cyg        & 20&46:25.54  & $+$40&06:59.4  & SiO & SIMBAD, Infrared\\
CRL2688        & 21&02:18.78  & $+$36&41:37.75 & HCN & APEX SHeFI\\
RV-Aqr         & 21&05:51.74  & $-$00&12:42.0  & HCN & SIMBAD, Infrared\\
EP-Aqr         & 21&46:31.866 & $-$02&12:45.71 & SiO & APEX SHeFI\\
pi1-Gru        & 22&22:44.241 & $-$45&56:52.83 & SiO & APEX SHeFI\\
S-Gru          & 22&26:05.474 & $-$48&26:18.76 & SiO & SIMBAD, Optical\\
RAFGL3068      & 23&19:12.61  & $+$17&11:33.1  & HCN & APEX SHeFI\\
W-Peg          & 23&19:50.501 & $+$26&16:43.66 & SiO & SIMBAD, Optical\\
R-Aqr          & 23&43:49.462 & $-$15&17:04.14 & SiO & SIMBAD, Optical
\end{longtable}
}
\end{appendix}

\end{document}